\documentclass[12pt]{article}
\usepackage{epsfig}

\pdfoutput=1

\def\beq{\begin{equation}}
\def\eeq#1{\label{#1}\end{equation}}
\def\eeqn{\end{equation}}

\def\beqa{\begin{eqnarray}}
\def\eeqa#1{\label{#1}\end{eqnarray}}
\def\eeqan{\end{eqnarray}}

\let\bar=\overbar

\def\Dslash{\not{\hbox{\kern-4pt $D$}}}
\def\dslash{\not{\hbox{\kern-2pt $\del$}}}

\def\msb{{\bar{\ssstyle M \kern -1pt S}}}

\usepackage{lineno}

\def\Title#1{\begin{center} {\Large {\bf #1} } \end{center}}

\begin{document}

\Title{Physics with electroweak penguins at LHCb}

\bigskip\bigskip


\begin{raggedright}  

{\it Michel De Cian \index{De Cian, M.}\\
on behalf of the LHCb collaboration \\
Physik-Institut, University of Z\"urich\\
Winterthurerstrasse 190\\
CH-8057 Z\"urich, SWITZERLAND}
\bigskip\bigskip
\end{raggedright}

\begin{raggedright}
\textit{Proceedings of CKM 2012, the 7th International Workshop on the CKM Unitarity Triangle, University of Cincinnati, USA, 28 September - 2 October 2012.}
\end{raggedright}


\section{Introduction}
Flavour changing neutral currents are only allowed via loop diagrams in the Standard Model (SM). Electroweak penguin processes are therefore sensitive probes for new physics, as physics beyond the Standard Model can enter via virtual particles at the same level as SM physics. The LHCb detector at the LHC~\cite{Alves:2008zz} with its forward geometry is ideally suited for the analysis of electroweak penguin processes in $B$ meson decays. All analyses are performed with 1 fb$^{-1}$ of collision data recorded at a centre-of-mass energy of 7 TeV in 2011 and constrain new physics models.

\section{Angular analysis and $\mathcal{CP}$-asymmetry in \\$B^{0} \rightarrow K^{* 0}\mu^{+}\mu^{-}$}
The decay $B^{0} \rightarrow K^{* 0}\mu^{+}\mu^{-}$ has a branching fraction of $\mathcal{B}(B^{0} \rightarrow K^{* 0}\mu^{+}\mu^{-}) = (1.05^{+0.16}_{-0.13}) \times 10^{-6}$~\cite{Beringer:1900zz} and can be fully described by four variables: the invariant mass of the dimuon system, $q^{2}$, and three angles, $\phi, \theta_{\ell}, \theta_{K}$. The angles are defined in Ref.~\cite{LHCb:2012a}. The analysis of $B^{0} \rightarrow K^{* 0}\mu^{+}\mu^{-}$ is performed in 6 bins of $q^{2}$ where the resonant regions of the $J/\psi$ and the $\psi(2S)$ are omitted~\cite{LHCb:2012a}. The decay is selected using a boosted decision tree~\cite{Breiman}; an event-by-event correction is applied to correct for experimental biases.

\begin{figure}[htb]
\begin{center}
\includegraphics[width=0.43\textwidth]{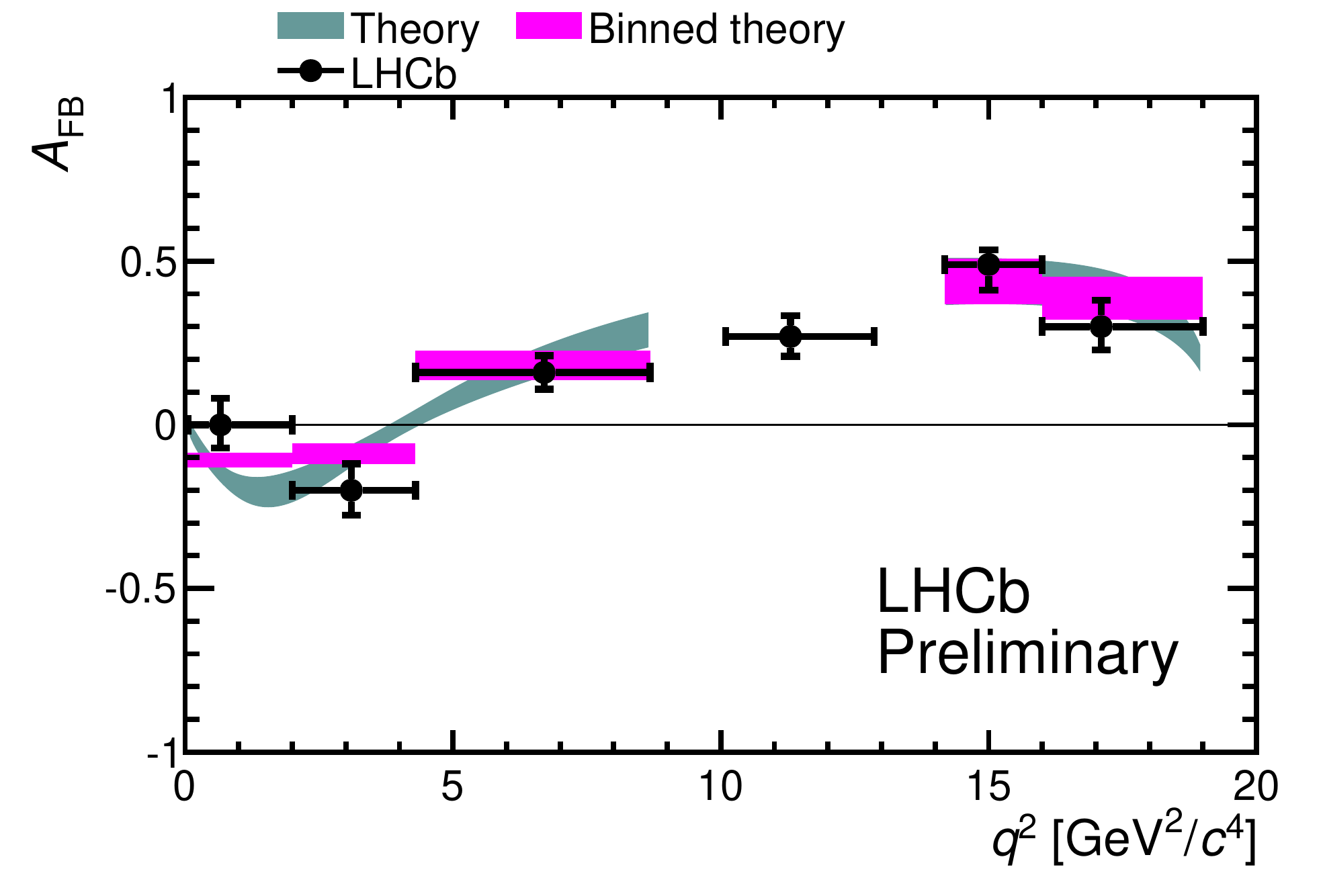}
\includegraphics[width=0.43\textwidth]{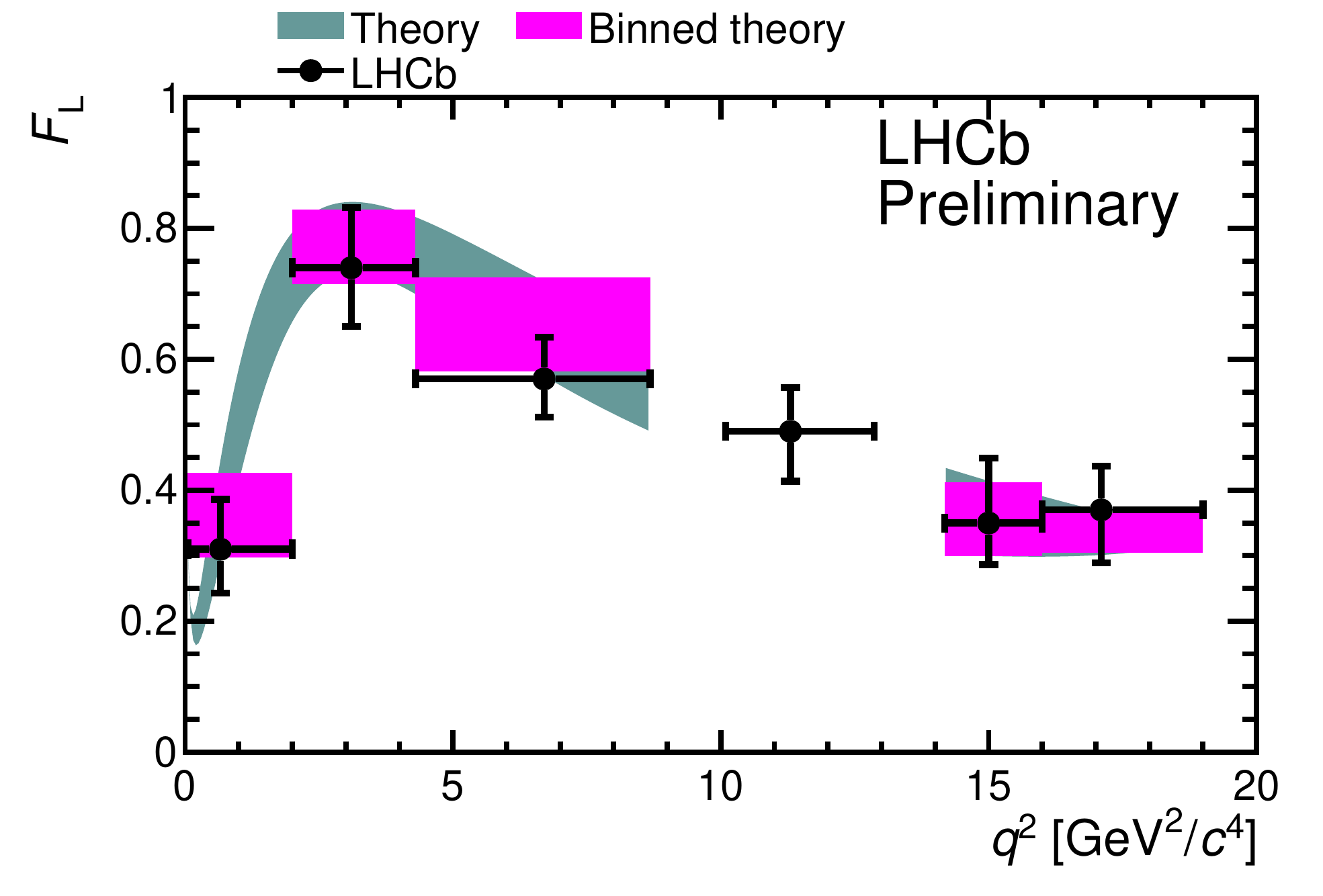}
\includegraphics[width=0.43\textwidth]{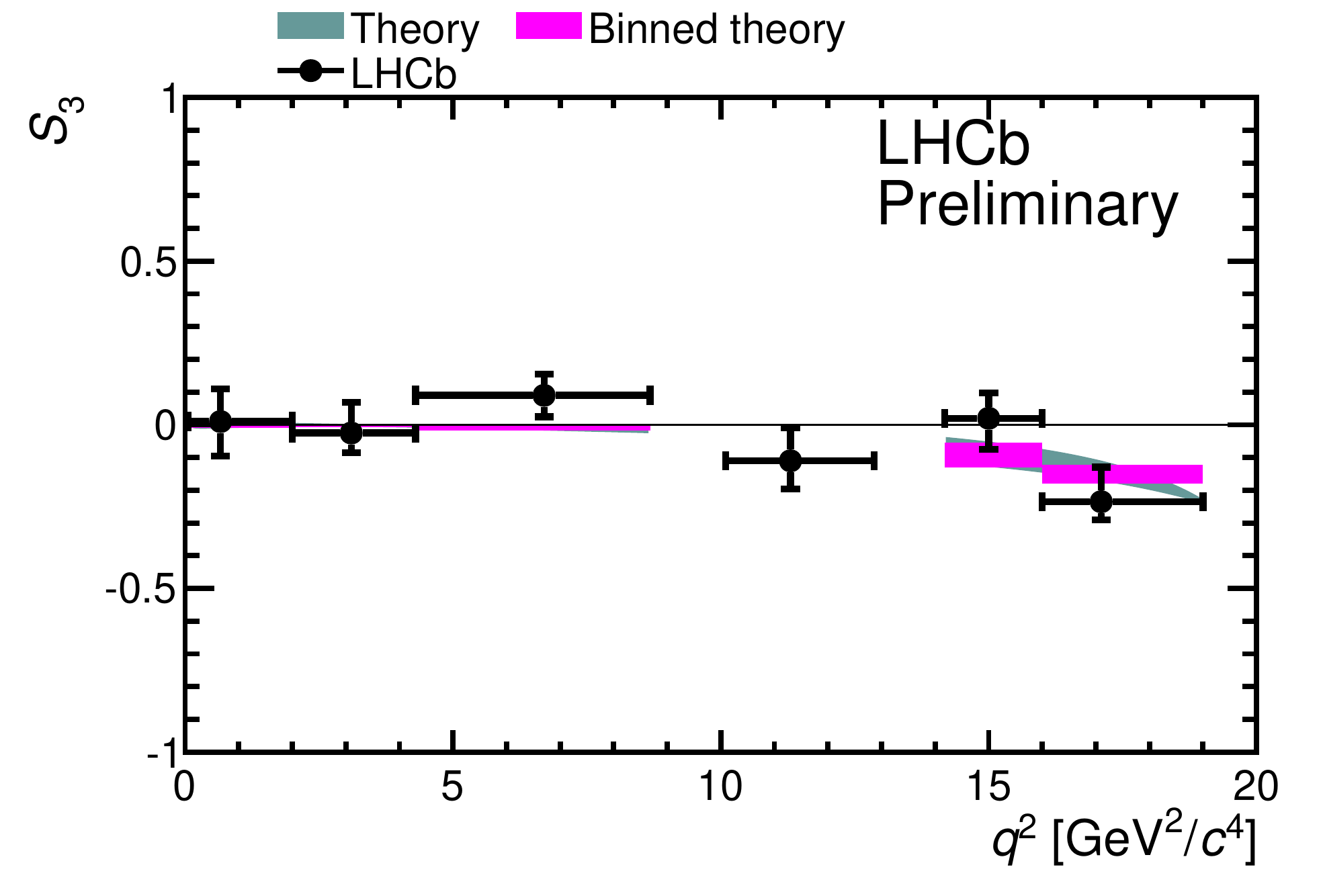}
\includegraphics[width=0.43\textwidth]{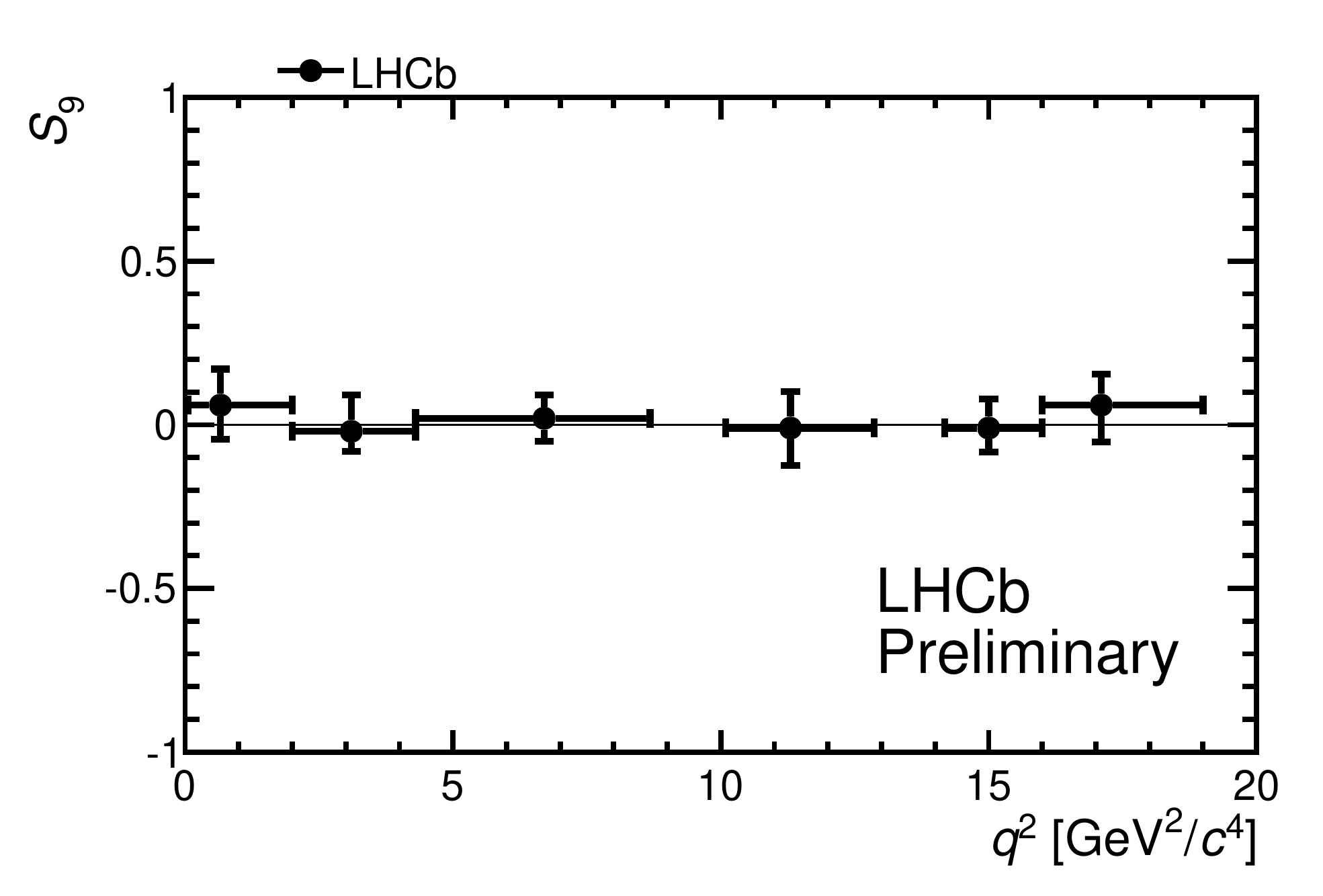}
\caption{$A_{FB}$, $F_{L}$, $S_{3}$ and $S_{9}$ as determind with an angular analysis in $B^{0} \rightarrow K^{* 0}\mu^{+}\mu^{-}$~\cite{LHCb:2012a}. The turquoise band shows the SM predictions, the pink bars the binned SM predictions. The SM prediction of $S_{9}$ is expected to be zero for the full $q^{2}$ range. The predictions are taken from Ref.~\cite{Bobeth:2011gi}. }
\label{fig:Kstmumu1}
\end{center}
\end{figure}

Four angular variables were examined: $A_{FB}$, the forward-backward asymmetry of the dimuon sytem; $F_{L}$, the longitudinal polarisation of the $K^{*0}$; $S_{3}$, a variable expressing the asymmetry between the $K^{*0}$ transverse and longitudinal polarisation; and $S_{9}$~\cite{Altmannshofer:2008dz}. The results together with the SM predictions are shown in Fig.~\ref{fig:Kstmumu1}. The zero-crossing point of $A_{FB}$ is an observable particularly sensitive to new physics contributions as the form-factor uncertainties cancel at first order in the theoretical prediction. The experimental value of the zero-crossing point is ($4.9^{+1.1}_{-1.3}$) GeV$^{2}/c^{4}$. It agrees well with the theoretical expectations, which are in the range [4.0 -- 4.3] GeV$^{2}$/c$^{4}$~\cite{Bobeth:2011nj}\cite{Beneke:2004dp}\cite{Ali:2006ew}. In addition to the angular variables, the differential branching fraction of $B^{0} \rightarrow K^{* 0}\mu^{+}\mu^{-}$ was measured as well.


\begin{figure}[htb]
\begin{center}
\includegraphics[width=0.43\textwidth]{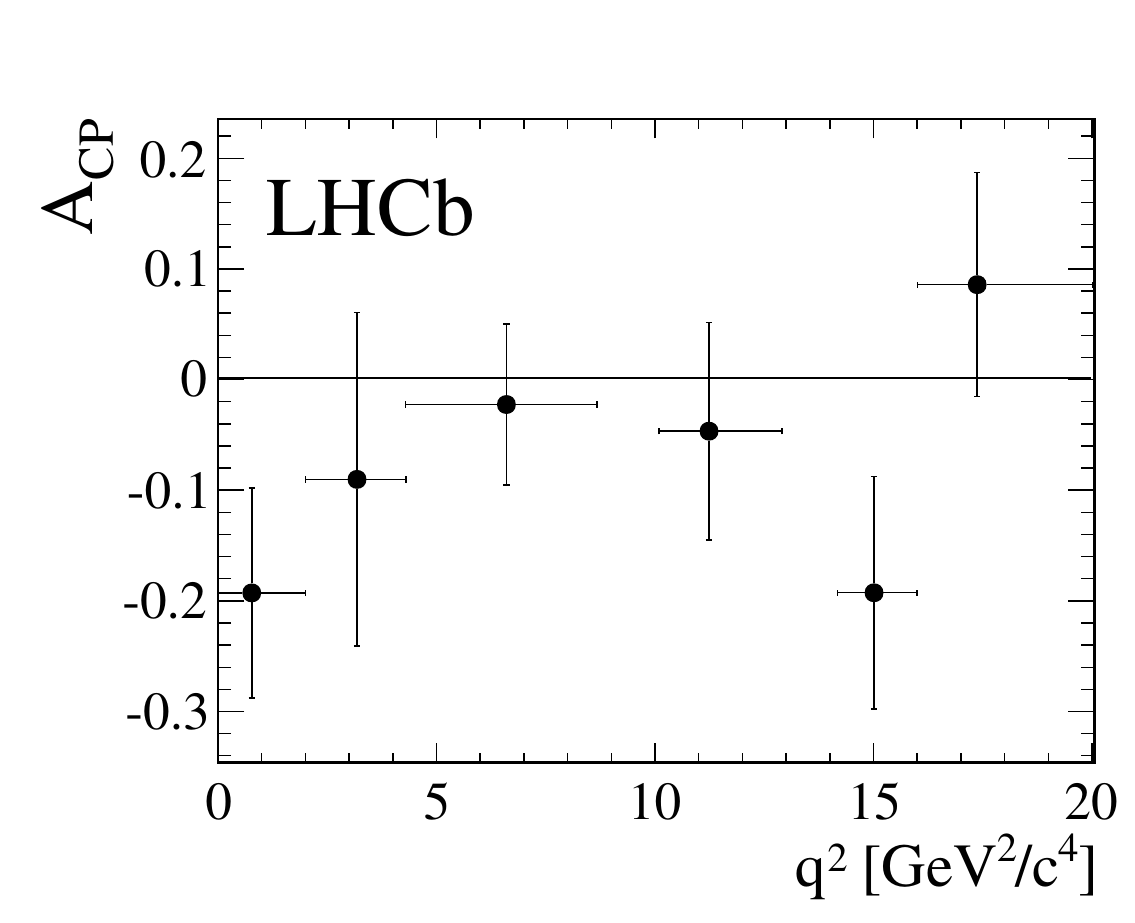}
\caption{{\cal CP}-Asymmetry in $B^{0} \rightarrow K^{* 0}\mu^{+}\mu^{-}$ as a function of the dimuon invariant mass squared~\cite{LHCb:2012kz}. The SM prediction is $\mathcal{O}(10^{-3})$ for the full range in $q^{2}$.}
\label{fig:KstmumuACP}
\end{center}
\end{figure}

The direct $\mathcal{CP}$-asymmetry in $B^{0} \rightarrow K^{* 0}\mu^{+}\mu^{-}$, $\mathcal{A_{CP}}$, is predicted to be $\mathcal{O}(10^{-3})$ in the SM and is defined as $\mathcal{A_{CP}} = \frac{\Gamma( \bar{B}^{0} \rightarrow \bar{K}^{* 0}\mu^{+}\mu^{-}) - \Gamma(B^{0} \rightarrow K^{* 0}\mu^{+}\mu^{-})}{\Gamma( \bar{B}^{0} \rightarrow \bar{K}^{* 0}\mu^{+}\mu^{-}) + \Gamma(B^{0} \rightarrow K^{* 0}\mu^{+}\mu^{-})}$. The theoretical prediction has a small uncertainty due to suppression of form factor uncertainties~\cite{Bobeth:2008ij}\cite{Altmannshofer:2008dz}. Models beyond the SM can enhance this value up to 15\%~\cite{Alok:2011gv}. The LHCb analysis uses the same event selection, correction for experimental effects and binning scheme in $q^{2}$ as the angular analysis of $B^{0} \rightarrow K^{* 0}\mu^{+}\mu^{-}$. Asymmetries due to detector effects are cancelled by taking an average with equal weights of the $\mathcal{CP}$-asymmetries measured in two independent data samples with opposite polarities of the LHCb dipole magnet. The production and interaction asymmetries are corrected for using the $B^{0}\rightarrow J/\psi K^{*0}$ decay mode as a control channel. Production asymmetries are also accounted for by considering $B^{0} \rightarrow J/\psi K^{* 0}$~\cite{LHCb:2012kz}.
The $\mathcal{CP}$-asymmetry in $B^{0} \rightarrow K^{* 0}\mu^{+}\mu^{-}$ in the 6 $q^{2}$ bins is shown in Fig.~\ref{fig:KstmumuACP}. The overall value, integrated over the $q^{2}$ bins, is $\mathcal{A_{CP}} = -0.072 \pm 0.040 (\textrm{stat}) \pm 0.005(\textrm{sys})$.

\section{Analysis of $B^{+} \rightarrow K^{+}\mu^{+}\mu^{-}$}
The analysis of $B^{+} \rightarrow K^{+}\mu^{+}\mu^{-}$ is similar to that of the $B^{0} \rightarrow K^{* 0}\mu^{+}\mu^{-}$ decay. The differential branching fraction is determined using $B^{+} \rightarrow J/\psi K^{+}$ as a normalisation channel. The differential branching fraction is determined in 7 bins of $q^{2}$. The integrated branching fraction, taking the region of the excluded charmonium resonances into account,  is $\mathcal{B}(B^{+}\rightarrow K^{+}\mu^{+}\mu^{-}) = (4.36 \pm 0.15(\mathrm{stat}) \pm 0.18(\mathrm{sys})) \times 10^{-7}$~\cite{Aaij:2012vr}. 

\begin{figure}[htb]
\begin{center}
\includegraphics[width=0.43\textwidth]{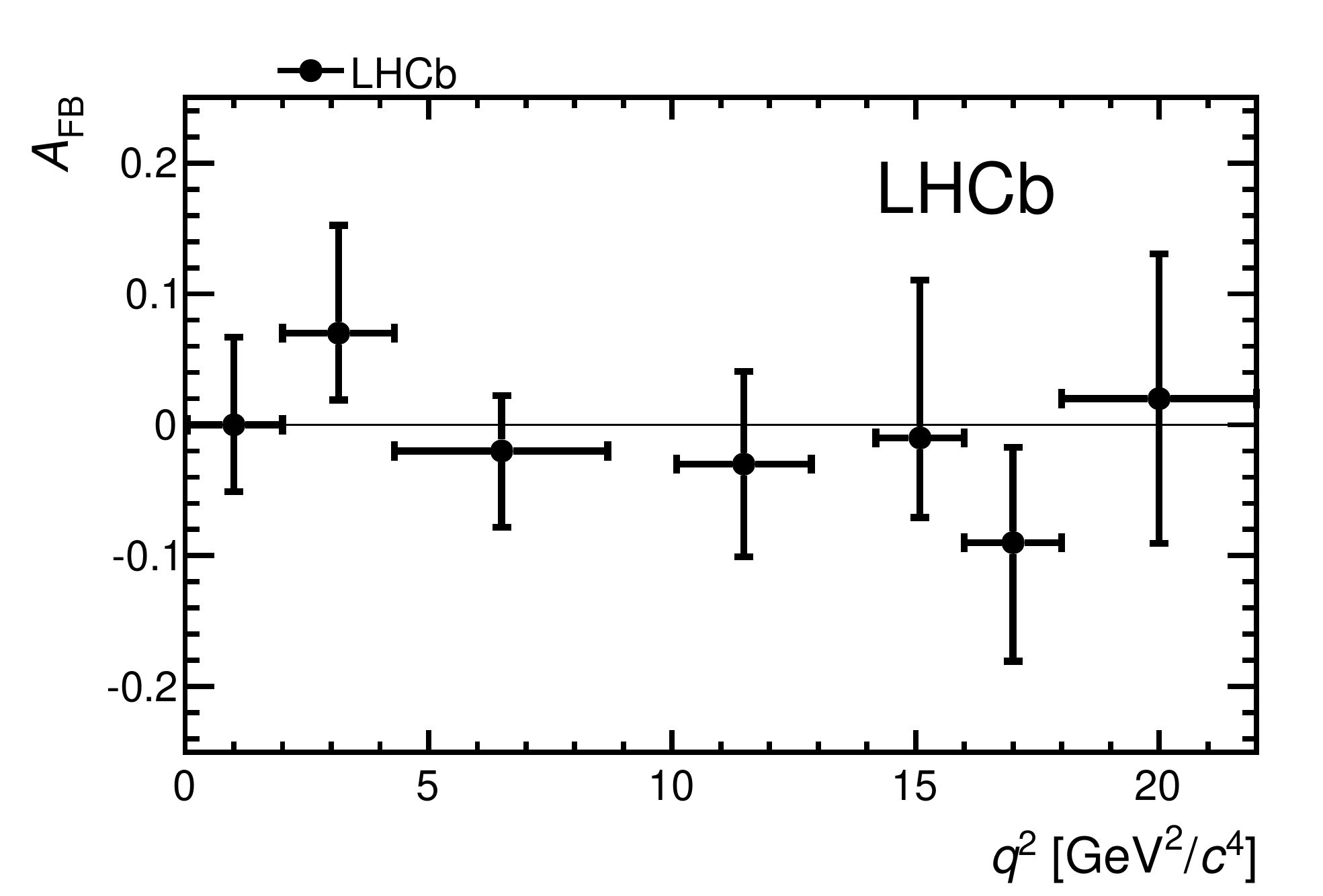}
\includegraphics[width=0.43\textwidth]{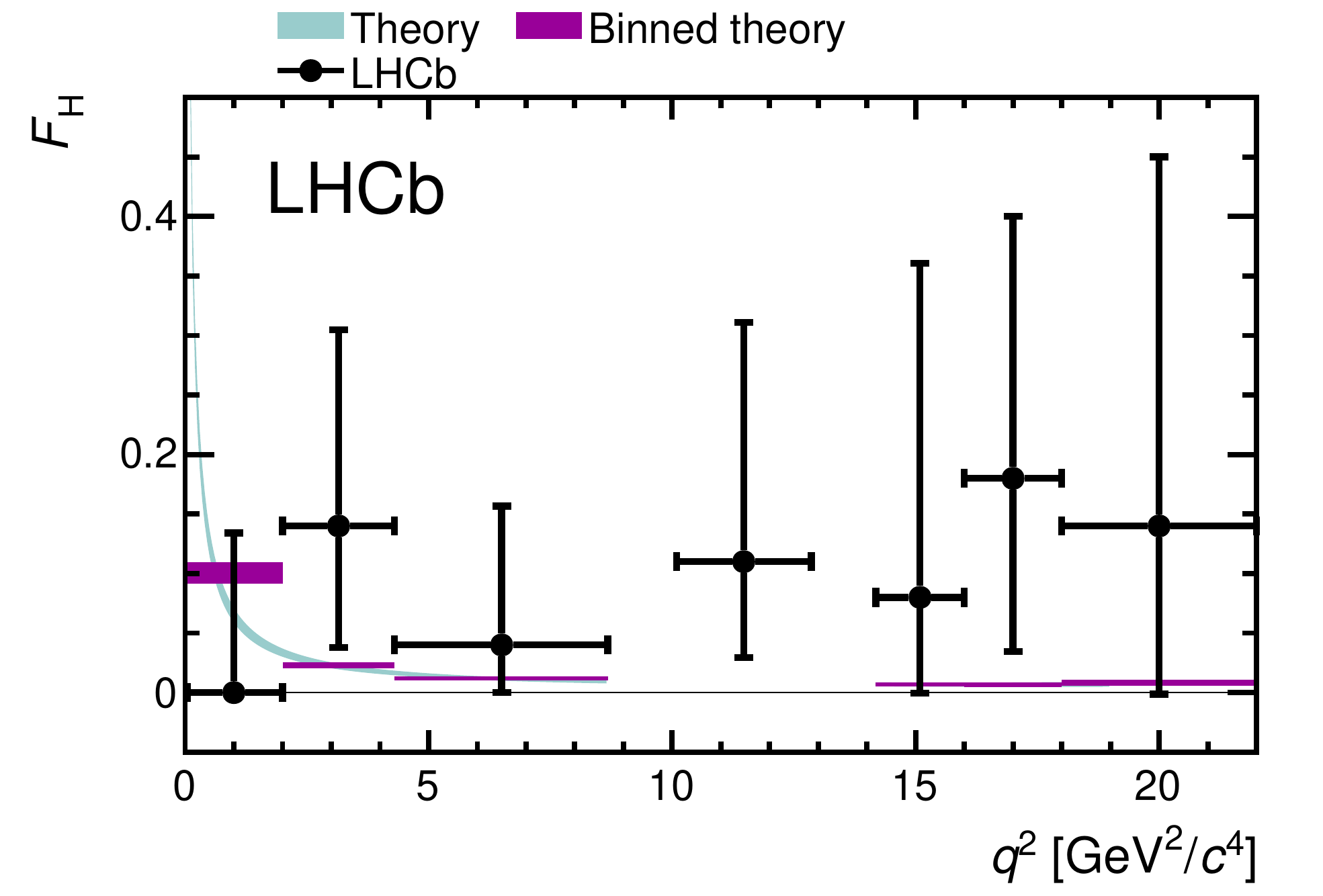}
\caption{The forward-backward asymmetry $A_{FB}$ and the flat parameter $F_{H}$ in $B^{+} \rightarrow K^{+}\mu^{+}\mu^{-}$ as a function of the dimuon invariant mass squared~\cite{Aaij:2012vr}. The theory predictions are taken from Ref.~\cite{Bobeth:2011nj} and \cite{Bobeth:2011gi}. $A_{FB}$ is expected to be negligible in the SM for the full $q^{2}$ range. }
\label{fig:Kmumu}
\end{center}
\end{figure}

The angular distribution of  $B^{+} \rightarrow K^{+}\mu^{+}\mu^{-}$ can be written as $\frac{1}{\Gamma}\frac{d\Gamma( B^{+} \rightarrow K^{+}\mu^{+}\mu^{-} )}{ d\cos \theta_{\ell}} = \frac{3}{4}(1-F_{H})(1 - \cos^{2}\theta_{\ell}) + \frac{1}{2}F_{H} + A_{FB}\cos\theta_{\ell}$ with $A_{FB}$ the forward-backward asymmetry and $F_{H}$ a flat parameter.  An event-by-event correction is applied to account for experimental effects. The resulting distributions of $A_{FB}$ and $F_{H}$ are shown in Fig.~\ref{fig:Kmumu}. They are in good agreement with the SM predictions.

\section{Isospin analysis of $B \rightarrow K^{(*)}\mu^{+}\mu^{-}$}
\begin{figure}[t]
\begin{center}
\includegraphics[width=0.43\textwidth]{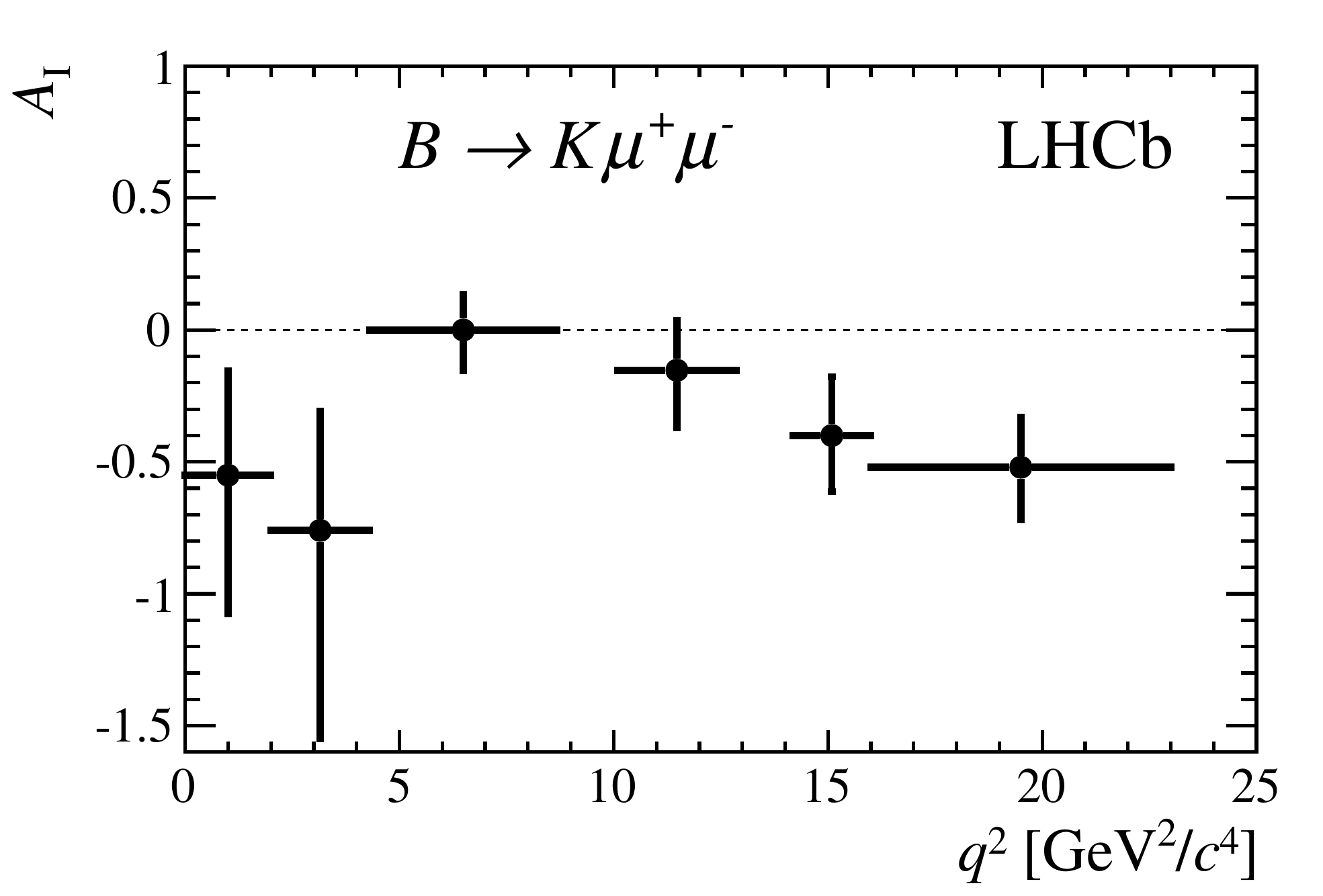}
\includegraphics[width=0.43\textwidth]{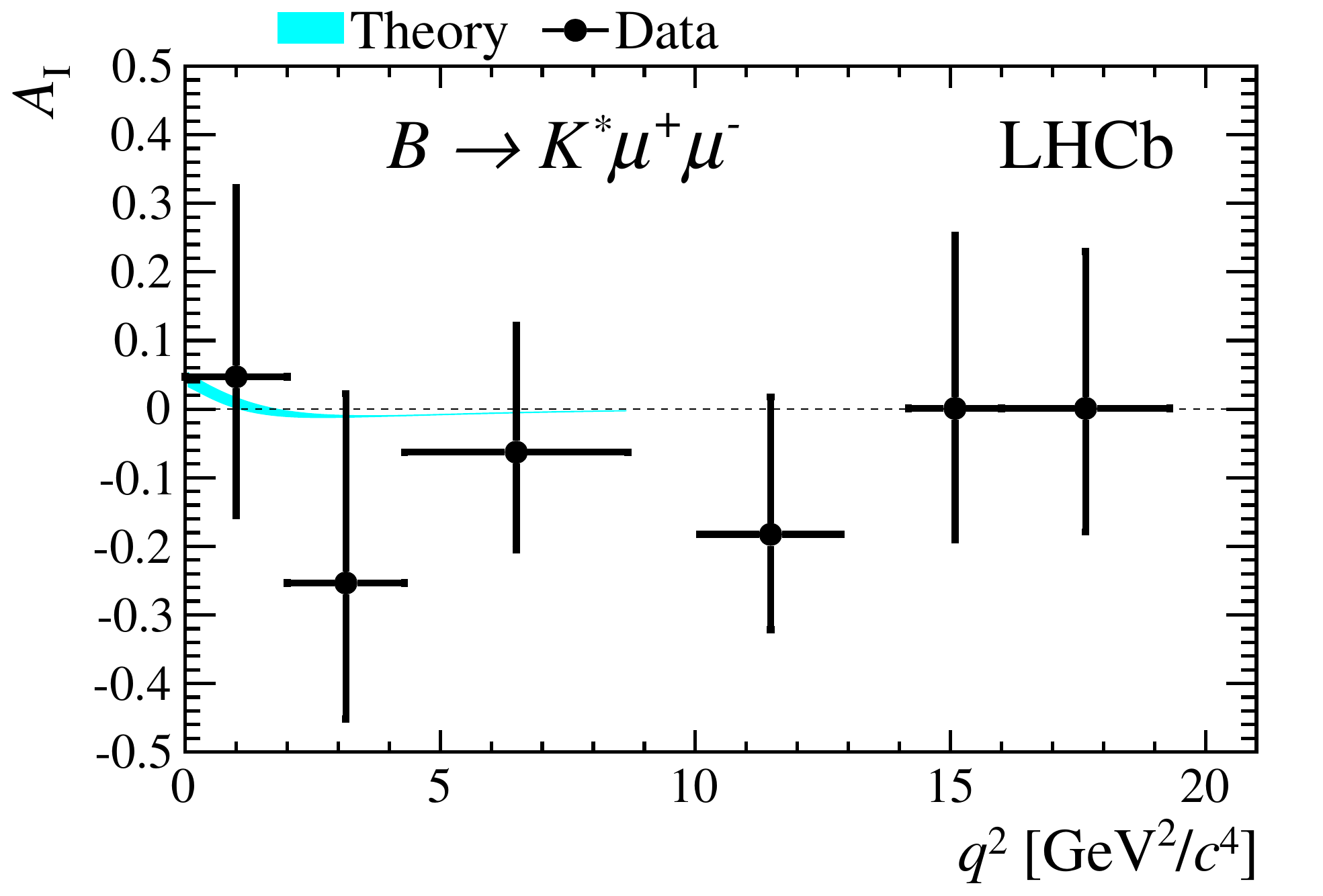}
\caption{Isospin asymmetry for $B\rightarrow K\mu^{+}\mu^{-}$ and  $B\rightarrow K^{*}\mu^{+}\mu^{-}$~\cite{Aaij:2012cq}. The SM predictions are taken from Ref.~\cite{Bobeth:2011nj} and \cite{Beneke:2001at}. }
\label{fig:Isospin}
\end{center}
\end{figure}

The so-called isospin asymmetry in the decays $B \rightarrow K^{(*)}\mu^{+}\mu^{-}$, $A_{I}$, is defined as $A_{I} = \frac{\Gamma ( B^{0} \rightarrow K^{(*)0}\mu^{+}\mu^{-} ) - \Gamma ( B^{+} \rightarrow K^{(*)+}\mu^{+}\mu^{-} )}{\Gamma ( B^{0} \rightarrow K^{(*)0}\mu^{+}\mu^{-}  ) + \Gamma ( B^{+} \rightarrow K^{(*)+}\mu^{+}\mu^{-} )}$ and is predicted to be very small in the SM~\cite{Feldmann:2002iw}. The LHCb analysis~\cite{Aaij:2012cq} consists of a measurement of four branching fractions: $B^{0}\rightarrow K_{S}^{0}\mu^{+}\mu^{-},\, B^{+}\rightarrow K^{+}\mu^{+}\mu^{-},\, B^{0}\rightarrow (K^{*0}\rightarrow K^{+}\pi^{-})\mu^{+}\mu^{-},\, B^{+}\rightarrow (K^{*+}\rightarrow K^{0}_{S}\pi^{+})\mu^{+}\mu^{-}$. $B \rightarrow J/\psi K^{(*)}$ is used as a normalisation channel. The isospin asymmetries are measured in 6 $q^{2}$ bins. For the $B \rightarrow K^{*}\mu^{+}\mu^{-}$ mode the isospin asymmetry agrees well with the SM prediction. For $B \rightarrow K\mu^{+}\mu^{-}$ the isospin asymmetry shows lower values compared with the expectation. This discrepancy is fully driven by the measured branching fraction of $B^{0}\rightarrow K^{0}\mu^{+}\mu^{-}$ which is too low compared to the SM prediction. Both isospin asymmetries are shown in Fig.~\ref{fig:Isospin}. When integrating over the full $q^{2}$ range, the deviation from the prediction is 4.4 standard deviations.

\section{Branching fraction measurement of $B^{+} \rightarrow \pi^{+} \mu^{+} \mu^{-}$}
The decay $B^{+} \rightarrow \pi^{+} \mu^{+} \mu^{-}$ is a $b \rightarrow d \ell \ell$ transition and therefore suppressed with respect to the $b\rightarrow s \ell \ell$ transitions by the ratio of the CKM matrix elements $\frac{|V_{td}|^{2}}{|V_{ts}|^{2}}$. The branching fraction is measured to be $\mathcal{B}(B^{+} \rightarrow \pi^{+} \mu^{+} \mu^{-}) = (2.3 \pm 0.6(\textrm{stat}) \pm 0.1(\textrm{sys})) \times 10^{-8}$~\cite{LHCb:2012de} which is in good agreement with the SM prediction of $(1.96 \pm 0.21) \times 10^{-8}$~\cite{Song:2008zzc}. Using the measurement of $B^{+}\rightarrow K^{+}\mu^{+}\mu^{-}$ the ratio $\frac{|V_{td}|}{|V_{ts}|}$ is determined to be $0.266 \pm 0.035 \textrm{ (stat)} \pm 0.007 \textrm{ (sys)}$ which is in good agreement with the results from radiative decays and $B^{0}$ and $B^{0}_{s}$ mixing~\cite{Amhis:2012bh}\cite{delAmoSanchez:2010ae}\cite{Abe:2005rj}\cite{Aubert:2006pu}.

\section{Conclusion}
Rare electroweak penguin decays are a very active research area in LHCb. With its rich decay structure the $B^{0} \rightarrow K^{* 0}\mu^{+}\mu^{-}$ decay is the golden channel for the study of $b \rightarrow s\ell\ell$ transitions.  Further decays like $B^{+} \rightarrow K^{+}\mu^{+}\mu^{-}$ can add additional information to search for physics beyond the SM. Up to now all measurements are compatible with the SM and set strong constraints on a broad range of new physics models.


\begin{thebibliography}{99}

\bibitem{Alves:2008zz}
  A.~A.~Alves, Jr. {\it et al.}  [LHCb Collaboration],
  JINST {\bf 3} (2008) S08005.

\bibitem{Beringer:1900zz}
  J.~Beringer {\it et al.}  [Particle Data Group Collaboration],
  Phys.\ Rev.\ D {\bf 86} (2012) 010001.

\bibitem{LHCb:2012a}
  R.~Aaij {\it et al.} [LHCb Collaboration],
  LHCb-CONF-2012-008

\bibitem{Breiman}
 Leo Breiman, Jerome Friedman, Charles J. Stone, and R. A. Olshen, Classification and Regression Trees.
 Wadsworth international group (1984)

\bibitem{Altmannshofer:2008dz}
  W.~Altmannshofer, P.~Ball, A.~Bharucha, A.~J.~Buras, D.~M.~Straub and M.~Wick,
  JHEP {\bf 0901} (2009) 019
  [arXiv:0811.1214 [hep-ph]].


\bibitem{Bobeth:2011nj}
  C.~Bobeth, G.~Hiller, D.~van Dyk and C.~Wacker,
  JHEP {\bf 1201} (2012) 107
  [arXiv:1111.2558 [hep-ph]].

\bibitem{Beneke:2004dp}
  M.~Beneke, T.~Feldmann and D.~Seidel,
  Eur.\ Phys.\ J.\ C {\bf 41} (2005) 173
  [hep-ph/0412400].

\bibitem{Ali:2006ew}
  A.~Ali, G.~Kramer and G.~-h.~Zhu,
  Eur.\ Phys.\ J.\ C {\bf 47} (2006) 625
  [hep-ph/0601034].

\bibitem{Bobeth:2008ij}
  C.~Bobeth, G.~Hiller and G.~Piranishvili,
  JHEP {\bf 0807} (2008) 106
  [arXiv:0805.2525 [hep-ph]].

\bibitem{Alok:2011gv}
  A.~K.~Alok, A.~Datta, A.~Dighe, M.~Duraisamy, D.~Ghosh and D.~London,
  JHEP {\bf 1111} (2011) 122
  [arXiv:1103.5344 [hep-ph]].



\bibitem{LHCb:2012kz}
  R.~Aaij {\it et al.}  [LHCb Collaboration],
  arXiv:1210.4492 [hep-ex].

\bibitem{Aaij:2012vr}
  R.~Aaij {\it et al.}  [LHCb Collaboration],
  arXiv:1209.4284 [hep-ex].

\bibitem{Feldmann:2002iw}
  T.~Feldmann and J.~Matias,
  JHEP {\bf 0301} (2003) 074
  [hep-ph/0212158].

\bibitem{Aaij:2012cq}
  R.~Aaij {\it et al.}  [LHCb Collaboration],
  JHEP {\bf 1207} (2012) 133
  [arXiv:1205.3422 [hep-ex]].

\bibitem{LHCb:2012de}
  R.~Aaij {\it et al.}  [LHCb Collaboration],
  arXiv:1210.2645 [hep-ex].

\bibitem{Song:2008zzc}
  H.~-Z.~Song, L.~-X.~Lu and G.~-R.~Lu,
  Commun.\ Theor.\ Phys.\  {\bf 50} (2008) 696.

\bibitem{Bobeth:2011gi}
  C.~Bobeth, G.~Hiller and D.~van Dyk,
  JHEP {\bf 1107} (2011) 067
  [arXiv:1105.0376 [hep-ph]].

\bibitem{Beneke:2001at}
  M.~Beneke, T.~Feldmann and D.~Seidel,
  Nucl.\ Phys.\ B {\bf 612} (2001) 25
  [hep-ph/0106067].






\bibitem{Amhis:2012bh}
  Y.~Amhis {\it et al.}  [Heavy Flavor Averaging Group Collaboration],
  arXiv:1207.1158 [hep-ex].

\bibitem{delAmoSanchez:2010ae}
  P.~del Amo Sanchez {\it et al.}  [BABAR Collaboration],
  Phys.\ Rev.\ D {\bf 82} (2010) 051101
  [arXiv:1005.4087 [hep-ex]].

\bibitem{Abe:2005rj}
  K.~Abe {\it et al.}  [Belle Collaboration],
  Phys.\ Rev.\ Lett.\  {\bf 96} (2006) 221601
  [hep-ex/0506079].

\bibitem{Aubert:2006pu}
  B.~Aubert {\it et al.}  [BABAR Collaboration],
  Phys.\ Rev.\ Lett.\  {\bf 98} (2007) 151802
  [hep-ex/0612017].

\end{thebibliography}
\end{document}